\documentclass[cameraready]{Interspeech}
\usepackage{tabularx}
\usepackage{booktabs}
\usepackage{multirow}
\usepackage{fix-cm}
\usepackage{amsmath}
\usepackage{booktabs}
\usepackage{graphicx}
\usepackage{multirow}
\usepackage{color}
\usepackage{enumitem}
\usepackage{threeparttable}
\usepackage{subcaption}
\usepackage{amssymb}

\title{Quality Adaptive Angular Margin Learning for Respiratory Sound Classification}

\usepackage[fontsize=9pt]{fontsize}

\author[affiliation={1}]{Yoon Tae}{Kim} 
\author[affiliation={1}]{Heejoon}{Koo}
\author[affiliation={1}]{Miika}{Toikkanen}
\author[affiliation={2,3}, orcid=0000-0003-0111-300X]{June-Woo}{Kim$^\dagger$}

\address{
    $^1$ RSC LAB, MODULABS, Republic of Korea \\
    $^2$ Department of Electronic Engineering, Wonkwang University, Republic of Korea \\
    $^3$ AI Convergence Research Institute, Wonkwang University, Republic of Korea
}
\email{dkimx3966@gmail.com, kaen2891@wku.ac.kr}


\usepackage{comment}

\begin{document}

\maketitle
\renewcommand{\thefootnote}{$\dagger$}
\footnotetext{Corresponding author.}
\begin{abstract}
We present a quality-adaptive angular-margin learning framework that improves feature generalization by enforcing intra-class compactness and inter-class separability. Our framework, titled QLung, introduces a no-reference audio quality margin derived from spectral entropy and root-mean-square energy, which adaptively scales angular margins based on recording quality. To this end, we propose a log-scaled angular margin that stabilizes training under severe class imbalance. We also use an angular classifier that normalizes features and class weights, ensuring margin penalties are applied consistently on the unit hypersphere. Our approach improves in-distribution performance on the ICBHI dataset by 2.46\% over the cross-entropy baseline, and most significantly, achieves the strongest out-of-distribution performance on the SPRSound dataset compared to prior state-of-the-art methods. Code is available at \textcolor{cyan}{\href{https://github.com/RSC-Toolkit/QLung}{https://github.com/RSC-Toolkit/QLung}}.
\end{abstract}

\section{Introduction}
\label{sec:introduction}

Respiratory sound classification (RSC) has attracted growing interest due to its potential to support the diagnosis of respiratory diseases. Recent progress has been driven by advances in model architectures ranging from CNNs to Transformers, which have enabled the learning of increasingly discriminative representations from respiratory recordings~\cite{ma2020lungrn+, yang2020adventitious, bae2023patch}. In addition to improvements, researchers have sought to exploit dataset information more thoroughly, for example by incorporating metadata to reduce bias and improve generalization~\cite{kim2024bts, toikkanen25_interspeech, moummad2023pretraining, kim2024stethoscope, 10902164, kim2025tri, koo2026empowering}

\begin{figure}[!t]
    \centering
    \includegraphics[width=0.7\linewidth]{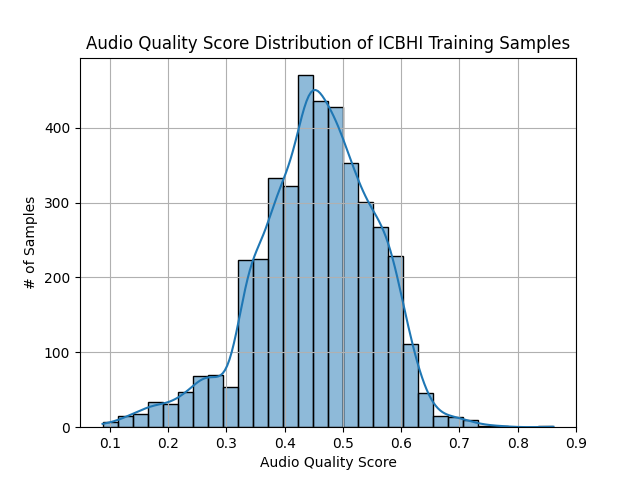}
    \caption{Audio quality score histogram of ICBHI training set.}
    \label{fig0}    
    \vspace{-5mm}
\end{figure}

Despite these advances, publicly available RSC datasets often contain variable-quality recordings or low-quality data.
Figure~\ref{fig0} shows the distribution of audio quality score based on spectral entropy and root-mean-square energy (RMS) in the ICBHI training set, with most samples concentrated in the mid-quality range (0.4--0.5), while the low-quality tail at the extreme end may yield degraded performance.
Identifying reliable samples is critical, as training on low-quality data risks amplifying noise rather than pathology, whereas emphasizing high-quality recordings can strengthen discriminative feature learning. Prior works have attempted to mitigate this issue through data augmentation~\cite{park19e_interspeech, kim2023adversarial, kim2024repaugment, ge2025lungmix, koo2026mitigating}, but unreliable samples still limit model robustness.
Consequently, such an issue undermines the model's ability to form clear and consistent decision boundaries, limiting generalization to real-world clinical data. 

Meanwhile, margin-based regularization has proven highly effective in domains such as face recognition~\cite{liu2017sphereface, deng2019arcface} or speaker verification~\cite{liu19f_interspeech}. ArcFace~\cite{deng2019arcface} imposes an additive angular margin between a feature vector and its target class weight, thereby enhancing discriminative power. This formulation is particularly appealing for RSC, where models must distinguish subtle boundaries between abnormal categories. Although \textit{crackle and wheeze} are acoustically distinct, their simultaneous occurrence in the \emph{both} class of RSC datasets introduces a fine-grained classification challenge, making it difficult to separate individual categories from their co-occurrence. 
Moreover, severe class imbalance leads to misclassification between abnormal and normal cases.
Margin-based learning is well-suited to address both the fine-grained separation of overlapping abnormal events and the imbalanced setting between normal and abnormal classes, as it enforces clearer decision boundaries. However, the application of angular margin learning to RSC has been largely unexplored. Motivated by this gap, we draw inspiration from additive angular margin regularization to learn more discriminative lung sound representations.

\begin{figure*}[!t]
    \centering
    \includegraphics[width=0.85\linewidth]{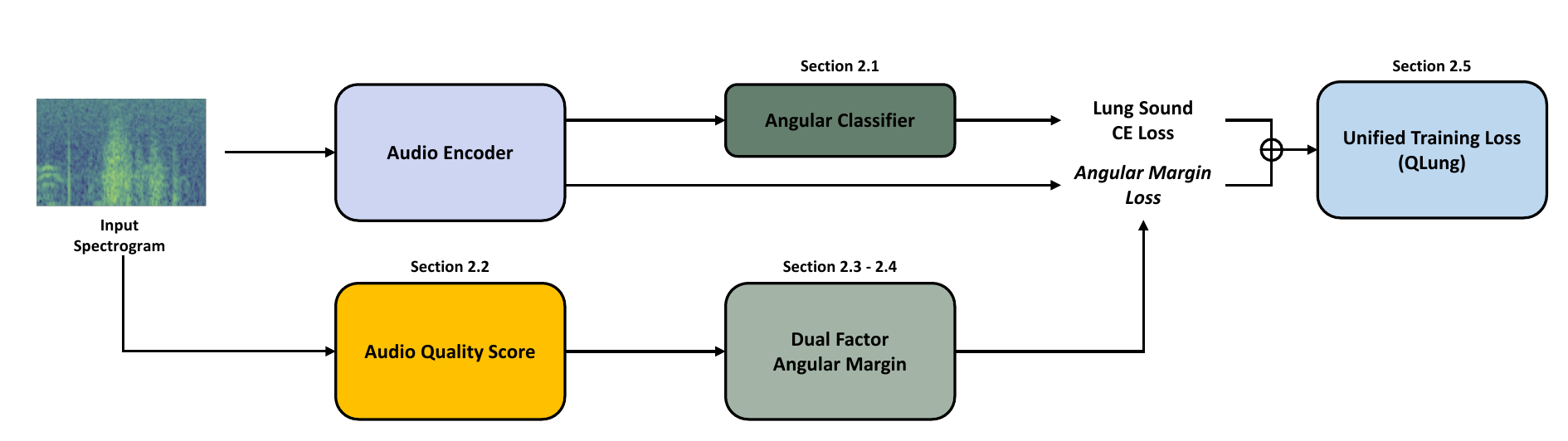}
    \caption{Illustration of proposed QLung framework.}
    \label{fig_first}    
\end{figure*}

We introduce QLung, which incorporates a quality-aware angular margin regularization scheme for RSC: 
\begin{itemize}[itemsep=0pt, topsep=0pt, parsep=0pt, partopsep=0pt, leftmargin=*]
    \item To the best of our knowledge, we are the first to apply the angular margin-based framework for RSC, addressing the chronic but under-explored problem of acoustically overlapping respiratory events and severe class imbalance by promoting more discriminative feature representations.

    \item We propose a unified dual-factor angular margin formulation that integrates $(i)$ a no-reference audio quality margin and $(ii)$ a log-scaled class-imbalance margin, enabling adaptive and stable margin learning under recording variability and severe imbalance.

    \item We design an angular classifier that normalizes features and class weights, constraining decisions to the unit hypersphere, ensuring the angular margin penalty is applied effectively.

    \item We obtain a 2.46\% improvement over AST fine-tuning~\cite{bae2023patch} on the ICBHI dataset~\cite{rocha2017alpha}, and achieve the best out-of-distribution (OOD) performance on the SPRSound dataset~\cite{zhang2022sprsound}, surpassing all prior state-of-the-art methods. 

\end{itemize}

\section{Methodology}
\label{sec:Methodology}
\subsection{Angular Classifier}
\label{sec:AngularClassifier}
Angular margin losses are designed to encourage features to cluster on the unit hypersphere in the directions of their class weight vectors~\cite{deng2019arcface, huang2020curricularface}. However, in a standard linear classifier, the target logit also depends on feature and weight norms, which can dilute purely angular clustering under angular margin regularization. Since loudness and recording quality largely perturb feature norms in respiratory sounds, angle-driven embeddings better separate subtle abnormal cues from normal breathing.

To sharpen the intended regularization signal, we introduce an angular classifier that normalizes both features and class weights and uses a fixed scale, so decisions depend purely on angular similarity. This strengthens the angular clustering effect of the margin penalty. Specifically, we formalize the angular classifier and its training objective as follows. Let 
\begin{equation}\label{eq:setup}
  W = [\,\mathbf{w}_1,\dots,\mathbf{w}_C\,] \in \mathbb{R}^{d \times C}, 
  \quad \mathbf{x} \in \mathbb{R}^{d}, 
  \quad y \in \{1,\dots,C\},
\end{equation}

\noindent where $C$ denotes the number of classes, $d$ denotes the feature dimensionality, $\mathbf{w}_k$ is the weight vector of class $k$, $\mathbf{x}$ is a feature vector, and $y$ is the ground-truth label. To remove the influence of vector magnitudes and make decisions purely on angular similarity, we apply L2 normalization to the features and the class weights, respectively:

\begin{equation}\label{eq:normalize}
  \hat{\mathbf{x}} = \frac{\mathbf{x}}{\lVert \mathbf{x} \rVert_2},
  \qquad
  \hat{\mathbf{w}}_k = \frac{\mathbf{w}}{\lVert \mathbf{w}_k \rVert_2},
  \qquad k=1,\dots,C.
\end{equation}

\noindent After normalization, inner products equal cosine similarities. With a fixed logit scale $s_a$, the class $k$ logit is
\begin{equation}\label{eq:logits}
z_k
=
s_a\,\hat{\mathbf{w}}_k^{\top}\hat{\mathbf{x}}
=
s_a\,\cos\theta_k,
\qquad
\theta_k \in \left[0,\pi\right],
\end{equation}
where $\theta_k$ is the angle between $\hat{\mathbf{w}}_k$ and $\hat{\mathbf{x}}$. 

The scale $s_a$ controls softmax sharpness and helps stabilize optimization, as is standard in angular margin losses. The predicted class probabilities:

\begin{equation}\label{eq:softmax} p_k = \frac{e^{z_k}}{\sum_{j=1}^{C} e^{z_j}}, \end{equation}
which define a categorical distribution over classes. For a single example with $y$, the cross-entropy (CE) loss is $L_{cls} \;=\; -\log p_{y}$.

\subsection{Audio Quality Margin}
\label{sec:AudioQualityScoreMargin}
Considering the variability in respiratory sound, we propose a no-reference \emph{Audio Quality Score} (AQS) in the range of $[0, 1]$, combining spectral entropy and RMS energy, which are widely used low-level acoustic features in speech and audio processing. High entropy values indicate noise-like signals, whereas low RMS energy corresponds to weak or poor-quality recordings. The AQS is given by:
\begin{equation}
\mathrm{AQS} = \text{clip}\left( 1 - \alpha H_{\mathrm{norm}} + \beta R_{\mathrm{norm}}, \, 0, 1 \right),
\end{equation}
where $H_{\mathrm{norm}}$ is normalized spectral entropy, $R_{\mathrm{norm}}$ is normalized RMS energy. 
We set $\alpha=0.7, \beta=0.3$ to balance noise sensitivity and signal strength. The $\text{clip}(\cdot,0,1)$ operator ensures that the score remains bounded between 0 and 1.  
To limit the peak AQS, we employ a scaling coefficient $\kappa$, defining the margin as $m_{q}=\kappa\,\mathrm{AQS}$. This quality margin keeps all recordings supervised, using larger margins for high-quality inputs to encourage stronger separation and lower margins for low-quality inputs to reduce noise overfitting, while modulating learning confidence rather than class targets.

\subsection{Log-scale Class Imbalance Margin}
\label{sec:Log-scaleClassImbalanceMargin}
To mitigate class imbalance and stabilize training, we introduce a log-scaled angular margin defined as the log of the inverse class frequency. A straightforward baseline is a scaled inverse-frequency margin. However, on severely imbalanced datasets like ICBHI, the margin can become excessively large in the tail, leading to unstable training. Accordingly, we adopt a logarithmic mapping that makes the margin vary smoothly and remains stable, rather than exploding in the tail. We formalize the proposed log-scaled angular class-imbalance margin below.

\noindent Let $n_y$ be the sample count of class $y$  and
$N=\sum_{k=1}^C n_k$ the total. 
To facilitate stable training, we incorporate class imbalance through the following \emph{log-scaled margin}:
\begin{equation}
m_{c_y} = s_c\,\big(-\log \pi_y\big),
\label{eq:log-imb-margin}
\end{equation}
where $\pi_y = \frac{n_y}{N}$ is class frequency.
Here $s_c$ is a scale chosen by the \emph{uniform anchor}, so that under a uniform
distribution ($\pi_y=\tfrac{1}{C}$), the margin equals the desired target $m_{\mathrm{target}}$:
\begin{equation}\label{eq:uniform-anchor} s_c = \frac{m_{\mathrm{target}}}{\log C}. \end{equation}

\begin{table*}[!t]
    \centering
    \caption{RSC results on the ICBHI datasets (60--40\% official). In the pretraining data column, IN, AS, and LA denote ImageNet~\cite{deng2009imagenet}, AudioSet~\cite{audioset}, and LAION-Audio-630K~\cite{wu2023large}, respectively. $*$ denotes the previous SOTA Score. \textbf{Best} and \underline{second-best} results.}
    
    \label{table3}
    \renewcommand{\arraystretch}{1}
    \addtolength{\tabcolsep}{8pt}
    \resizebox{\linewidth}{!}{
    \begin{tabular}{lccc|lll}
    \toprule
    Method & Backbone & Pretraining Data & Venue & $S_p$\,(\%) & $S_e$\,(\%) & \textbf{Score}\,(\%) \\
    \hline \midrule

    SE+SA \cite{yang2020adventitious} & ResNet18 & - & \textit{INTERSPEECH`20} & {81.25} & 17.84 & 49.55 \\

    LungRN+NL \cite{ma2020lungrn+} & ResNet-NL & - & \textit{INTERSPEECH`20} & 63.20 & 41.32 & 52.26 \\
    

    
    Ren \textit{et al.} \cite{ren2022prototype} & CNN8-Pt & - & \textit{ICASSP`22} & 72.96 & 27.78 & 50.37 \\
    
    Wang \textit{et al.} \cite{wang2022domain} (Splice) & ResNeSt & IN & \textit{ICASSP`22} & 70.40 & 40.20 & 55.30 \\


    
    
    
    
    Bae \textit{et al.} \cite{bae2023patch}\, (Fine-tuning)  & AST & IN\,+\,AS & \textit{INTERSPEECH`23} & $\text{77.14}$ & $\text{41.97}$ & $\text{59.55}$ \\
    
    Bae \textit{et al.} \cite{bae2023patch}\, (Patch-Mix CL) & AST & IN\,+\,AS & \textit{INTERSPEECH`23} & $\text{81.66}$ & $\text{{43.07}}$ & $\text{62.37}$ \\
    
    
    Kim \textit{et al.} \cite{kim2024stethoscope}\, (SG-SCL) & AST & IN\,+\,AS & \textit{ICASSP`24} & $\text{{79.87}}$ & $\text{{43.55}}$ & $\text{{61.71}}$ \\


    Xiao \textit{et al.} 
    \cite{xiao24_interspeech} (LungAdapter) & \text{AST} & IN\,+\,AS & \textit{INTERSPEECH`24} & $\text{{80.43}}$ & $\text{44.37}$ & $\text{{62.40}}$ \\

    Kim \textit{et al.} 
    \cite{kim2024bts} (Audio-CLAP) & \text{CLAP} & LA & \textit{INTERSPEECH`24} & $\text{{80.85}}$ & $\text{{44.67}}$ & $\text{{62.56}}$ \\
    
    Kim \textit{et al.} 
    \cite{kim2024bts} (BTS) & CLAP & LA & \textit{INTERSPEECH`24} & $\text{81.40}$ & $\textbf{{45.67}}$ & $\textbf{{63.54}}^\textbf{*}$ \\

    Ge \textit{et al.} 
    \cite{ge2025lungmix} (Lungmix) & \text{AST} & IN\,+\,AS & \textit{ICASSP`25} & $\text{{--}}$ & $\text{--}$ & $\text{{58.53}}$ \\

    \midrule
    
    \textbf{QLung on AST [ours]} & \text{AST} & IN\,+\,AS & \textit{--} & $\underline{{81.90}}_{\pm 5.38}$ & $\text{42.12}_{\pm 3.94}$ & $\text{62.01}_{\pm 1.18}$ \\

    \textbf{QLung on Audio-CLAP [ours]} & \text{CLAP} & LA & \textit{--} & $\text{\textbf{81.98}}_{\pm 3.82}$ & $\underline{44.81}_{\pm 3.47}$ & $\underline{63.39}_{\pm 0.40}$ \\


    
    \bottomrule
    \end{tabular}}
    \vspace{-3mm}
\end{table*}
\noindent With $x_y=-\log\pi_y$, the margin is linear in log-frequency:
\begin{equation}
m_{c_y}=s_c x_y=s_c(\log N-\log n_y).    
\end{equation}
Thus, it is affine in $\log n_y$ with slope $-s_c$; A factor $\rho$ decrease in frequency increases the margin by $s_c \log \rho$. This ensures smooth margin growth, emphasizing minority classes without over-penalizing majority classes, and stabilizing the trade-off between head and tail classes.

\subsection{Dual Factor Angular Margin Regularization (DFAM)}
\label{sec:DualFactorAngularMarginRegularization}
We propose DFAM, where the margin is formed as a weighted average of an audio quality term and a log-scaled class imbalance term. 
Specifically, we define the composite margin of the audio-quality margin $m_q$ and the class-imbalance margin $m_c$
via the relative weighting $\gamma$ as $m_{d} = \gamma\, m_{q} + (1-\gamma) m_{c}$.
The relative weight controls the tradeoff between quality adaptive regularization and class imbalance compensation in the composite margin.
Building on this, we apply the computed margin as a penalty on the target angle and compute the additive angular margin loss as:
\begin{equation}\label{eq:dfam}
\begin{aligned}
L_{\mathrm{DFAM}}
&=
-\log
\left(
\frac{
e^{s_d \cos(\theta_y + m_d)}
}{
e^{s_d \cos(\theta_y + m_d)}
+
\sum_{j \neq y} e^{s_d \cos(\theta_j)}
}
\right),
\end{aligned}
\end{equation}
where $s_d$ is the logit scale.

\subsection{QLung: Unified Training Objective}
\label{sec:QLung:UnifiedTrainingObjective}
Jointly optimizing the CE loss and the DFAM regularization sharpens decision boundaries while simultaneously promoting angular clustering of features. The final training objective is:

\begin{equation}
L_{\text{total}} = L_{\mathrm{cls}} + \lambda\,L_{\mathrm{DFAM}},
\end{equation}
where $\lambda$ denotes the weight of the DFAM regularization term.

\section{Experiments}
\label{sec:experiments}

\subsection{Experimental Setup}
\label{sec:experimental_setup}

\noindent\textbf{ICBHI Dataset.} 
We evaluated our method on the ICBHI 2017 respiratory sound dataset~\cite{rocha2017alpha}, a widely used benchmark for RSC. The dataset comprises approximately 5.5 hours of recordings, segmented into 6,898 respiratory cycles. Following the official protocol, the data are split into training (60\%) and test (40\%) subsets at the cycle level, with no patient overlap between sets. This yields 4,142 training and 2,756 test samples distributed across four respiratory sound categories: \emph{normal, crackle, wheeze, both}. In total, the dataset contains 3,642 \emph{normal} (52.8\%), 1,864 \emph{crackle} (27.0\%), 886 \emph{wheeze} (12.9\%), and 506 cycles with \emph{both} (wheeze+crackle) (7.3\%).

\noindent\textbf{SPRSound Dataset.} 
To further verify the generalization of our method against distribution shifts, we used the SPRSound dataset~\cite{zhang2022sprsound}, which provides seven respiratory sound classes. For compatibility with the ICBHI, we merged \emph{coarse crackle} and \emph{fine crackle} into a single \emph{crackle} class, as well as \emph{stridor} and \emph{rhonchi} into the \emph{wheeze} category. Finally, the overall distribution consists of 6,199 \emph{normal} (76.7\%), 1,044 \emph{crackle} (12.9\%), 811 \emph{wheeze} (10.0\%), and 31 cycles with \emph{both} class (0.4\%).

\noindent\textbf{Training Details.} 
For all experiments, we followed the data pre-processing described in prior studies~\cite{bae2023patch, kim2024bts, kim2024stethoscope, kim2024repaugment}, standardizing each respiratory cycle to 8 seconds. 
The AST~\cite{gong2021ast} model was used as our baseline, while additional experiments with Audio-CLAP~\cite{kim2024bts, wu2023large} were conducted to demonstrate the cross-architecture applicability of our approach. 
For AST, we extracted 128-dimensional log-Mel filterbanks using a 25 ms window and 10 ms hop size, with SpecAugment~\cite{park19e_interspeech} applied for robustness. For Audio-CLAP, the LAION-CLAP-630K~\cite{wu2023large} tokenizer without augmentation was used. All models were trained with the Adam~\cite{kingma2014adam} optimizer using a learning rate of 5e--5, and a batch size of 8 for 50 epochs. 
For QLung, we set $\lambda=0.4$, $\gamma=0.5$, $m_{\mathrm{target}}=0.2$, $s_a=37$, $s_d=15$, and $\kappa=0.5$, as chosen based on the sensitivity study in Figure~\ref{fig3}.

\noindent\textbf{Metrics.} 
We evaluated the performance using the official ICBHI metrics specificity ($S_p$), sensitivity ($S_e$), and their arithmetic mean, referred to as the Score~\cite{ bae2023patch, kim2024bts, kim2024stethoscope, kim2025tri, kim2024repaugment, rocha2017alpha, kim2025adaptive, koo2026empowering}. 
$S_p$ denotes the proportion of normal cases correctly classified, while $S_e$ indicates the proportion of correctly classified abnormal cases. All results are reported as the mean and standard deviation over five seeds (1--5).

\subsection{Experimental Results}
\label{sec:experimental_results}

\subsubsection{Overall ICBHI Dataset Results}
\begin{table*}[!t]
\centering
\caption{Comparative studies on the ICBHI (in-distribution) and SPRSound (out-of-distribution) datasets. \textbf{Best} results are highlighted.}
\scriptsize
\resizebox{0.8\linewidth}{!}{%
\begin{tabular}{lllllll}
\toprule
\multirow{2}{*}{Methods} & \multicolumn{3}{c}{ICBHI (In-Distribution)} & \multicolumn{3}{c}{SPRSound (Out-of-Distribution)} \\
\cmidrule(lr){2-4}\cmidrule(lr){5-7}
& $S_p$ (\%) & $S_e$ (\%) & \textbf{Score} (\%) & $S_p$ (\%) & $S_e$ (\%) & \textbf{Score} (\%) \\
\midrule
Bae et al. \cite{bae2023patch} (Patch-Mix CL)               & 81.66 & 43.07 & 62.37 & 62.69 & 39.33 & 51.01 \\
Kim et al. \cite{kim2024stethoscope} (SG-SCL)               & 79.87 & 43.55 & 61.71 & \text{81.06} & 22.62 & 51.84 \\
Kim et al. \cite{kim2024bts} (Audio-CLAP)              & 80.85 & 44.67 & 62.56 & 70.67 & \text{41.90} & 56.29 \\
Kim et al. \cite{kim2024bts} (BTS)        & 81.40 & \textbf{45.67} & \textbf{63.54} & 67.50 & 39.33 & 53.42 \\
\textbf{Ours (QLung on AST)} & $\text{81.90}_{\pm 5.38}$ & $\text{42.12}_{\pm 3.94}$ & $\text{62.01}_{\pm 1.18}$ & $\textbf{{82.48}}_{\pm 8.35}$ & $\text{33.99}_{\pm 8.21}$ & $\text{{58.23}}_{\pm 3.83}$ \\
\textbf{Ours (QLung on Audio-CLAP)} & $\textbf{{81.98}}_{\pm 3.82}$ & $\text{44.81}_{\pm 3.47}$ & $\text{63.39}_{\pm 0.40}$ & $\text{{74.71}}_{\pm 4.04}$ & $\textbf{44.88}_{\pm 2.98}$ & $\textbf{59.80}_{\pm 3.51}$ \\

\bottomrule
\end{tabular}
}
\label{table4}
\vspace{-3mm}
\end{table*}

Table~\ref{table3} presents a comprehensive comparison of RSC results on the ICBHI dataset under the official 60--40\% train--test split. Early CNN- and ResNet-based models showed limited performance, with Scores below 55\%~\cite{ma2020lungrn+, yang2020adventitious, ren2022prototype, wang2022domain}. The introduction of AST~\cite{gong2021ast} and its variants~\cite{bae2023patch, xiao24_interspeech} improved generalization, raising Scores to the range of 59--62\%. More recently, the adoption of metadata-guided learning strategies~\cite{kim2024bts, kim2024stethoscope} has further advanced the state-of-the-art, with BTS reporting the previous best Score of 63.54\%.
Applying QLung to the AST and Audio-CLAP backbones showed highly competitive results and surpassed their original performance without relying on additional augmentation strategies such as Patch-Mix~\cite{bae2023patch}, Lungmix~\cite{ge2025lungmix}, or additional LoRA~\cite{xiao24_interspeech} training. For both AST and Audio-CLAP, QLung improved the Score from 59.55\% to 62.01\% and from 62.56\% to 63.39\%, respectively. These results emphasize that QLung generalizes across architectures while remaining competitive against the latest state-of-the-art methods.

\subsubsection{Comparative In/Out-of-Distribution Studies}

Table~\ref{table4} reports a comparative evaluation of QLung with recent high-performing methods on both the in-distribution ICBHI dataset and the SPRSound~\cite{zhang2022sprsound} (OOD) dataset. While QLung achieved competitive performance on ICBHI, its advantage was most evident in the OOD evaluation on SPRSound. Whereas prior methods suffered from significant performance degradation, QLung attained the best result with a Score of 59.80\%, clearly outperforming BTS and Audio-CLAP. These results emphasize that, beyond in-distribution performance, the proposed QLung provides superior generalization to OOD datasets, demonstrating itself as a reliable framework for real-world clinical scenarios where recording conditions are noisy, heterogeneous, and substantial distribution shifts.

\subsubsection{Ablation Study: Impact of Proposed Components}
\begin{table}[!t]
\centering
\caption{
ICBHI results with AST backbone. Baseline: CE; components added incrementally.
\textbf{Best} result is highlighted.}
\label{table1}
\resizebox{\columnwidth}{!}{%
\begin{tabular}{lccc}
\toprule
\textbf{Method} & \textbf{$S_p$} & \textbf{$S_e$} & \textbf{Score} \\
\midrule
AST CE~\cite{bae2023patch} & $77.14_{\,\pm 3.35}$ & $41.97_{\,\pm 2.84}$ & $59.55_{\,\pm 1.92}$ \\
+ Fixed angular margin & $75.65_{\,\pm 6.13}$ & $\text{\textbf{43.77}}_{\,\pm 6.09}$ & $59.72_{\,\pm 0.89}$ \\
+ Audio quality & $80.99_{\,\pm 3.69}$ & $39.95_{\,\pm 5.32}$ & $60.47_{\,\pm 0.96}$ \\
+ Class imbalance & $79.16_{\,\pm 6.17}$ & $41.95_{\,\pm 6.02}$ & $60.56_{\,\pm 0.83}$ \\
+ Angular classifier (QLung) & $\text{\textbf{81.90}}_{\,\pm 5.38}$ & $42.12_{\,\pm 3.94}$ & $\text{\textbf{62.01}}_{\,\pm 1.18}$ \\
\bottomrule
\end{tabular}%
}
\vspace{-3mm}
\end{table}
Table~\ref{table1} presents the results of our ablation study, where we progressively integrated our proposed components into the AST baseline trained with CE loss~\cite{bae2023patch}. Adding the angular margin clearly improved abnormal sound classification performance, while the audio quality margin traded off sensitivity for higher specificity, leading to an overall Score improvement. Class-imbalance correction provided a modest additional gain to the proposed audio quality margin. The QLung achieved the best overall performance with a Score of 62.01\%, yielding a $+2.46$\% absolute improvement over the baseline. Our results validate the complementary impact of each component and the effectiveness of QLung as a robust framework for RSC tasks.

\begin{figure}[!t]
    \centering
    \begin{subfigure}{.5\linewidth}
      \centering
      \includegraphics[width=0.92\linewidth]{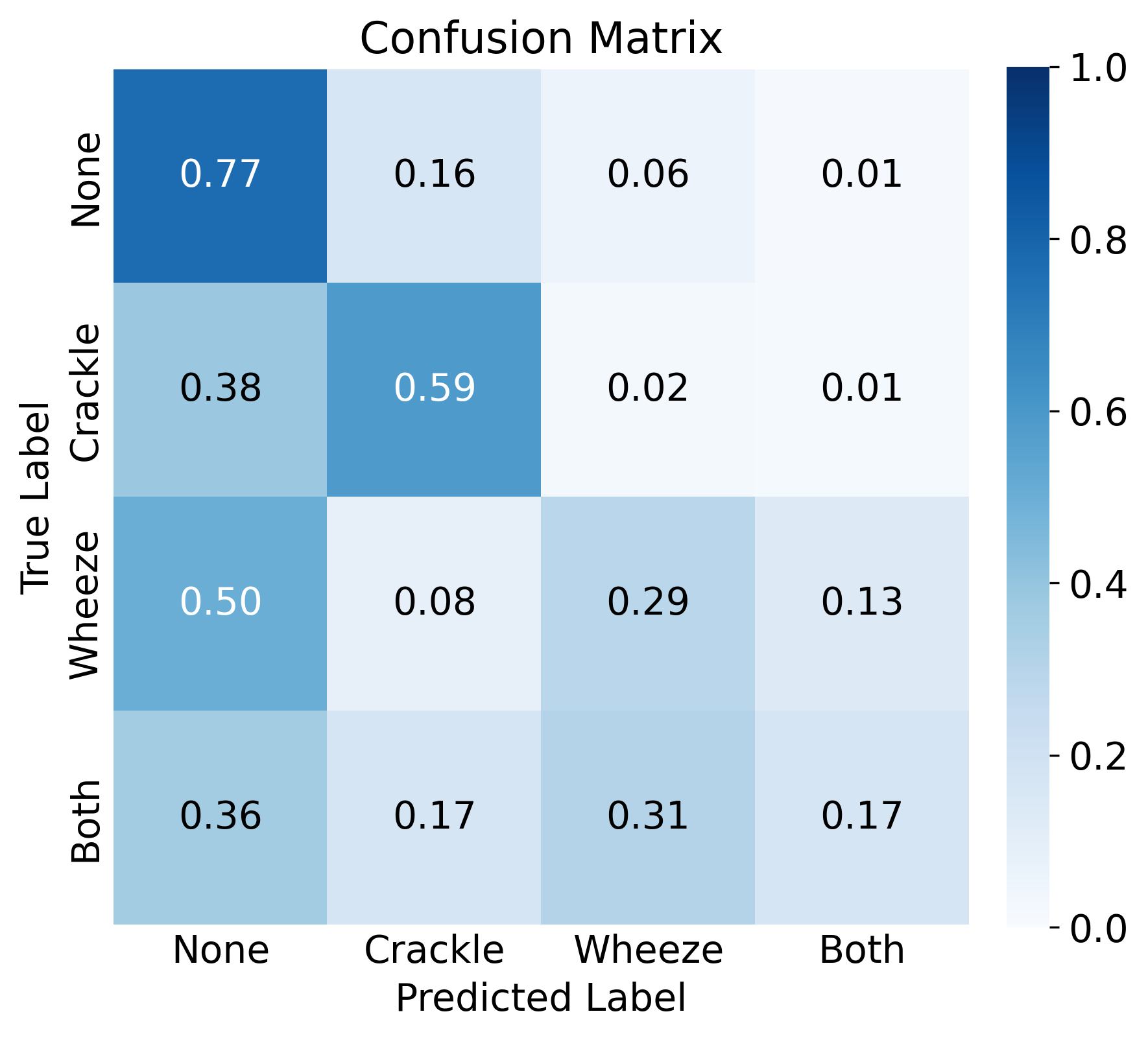}
      \caption{CE on AST}
      \label{fig:sfig1}
    \end{subfigure}%
    \begin{subfigure}{.5\linewidth}
      \centering
      \includegraphics[width=0.92\linewidth]{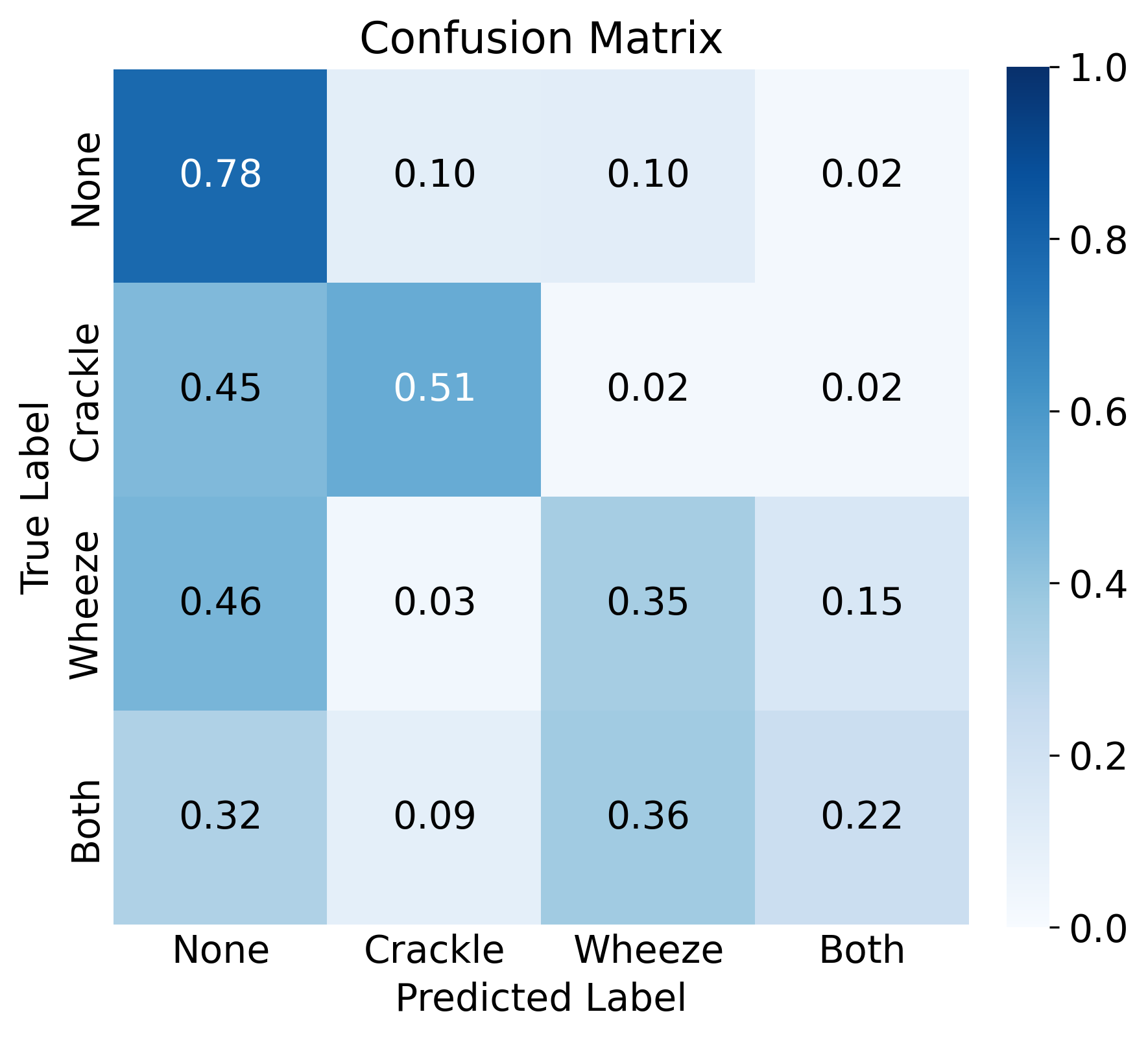}
      \caption{QLung on AST}
      \label{fig:sfig2}
    \end{subfigure}
    \caption{Confusion matrix results of CE and QLung on AST.}
    \label{fig1} 
    \vspace{-5mm}
\end{figure}

\subsubsection{Qualitative Analysis}
\textbf{Confusion Matrix Analysis.} 
Figure~\ref{fig1} shows that QLung improved classification performance compared to CE on AST, with higher accuracy for normal (+1\%), wheeze (+6\%), and \emph{both} (+5\%) cases, while reducing accuracy for crackle (-8\%). Although crackle accuracy decreases, QLung substantially reduces false crackle predictions for normal cases (-6\%) and alleviates crackle confusion for the both class (-8\%), while improving recognition of wheeze and both.  
Collectively, these changes indicate that QLung better recognizes clinically relevant abnormal classes while reducing false alarms even in OOD setting.

\begin{figure}[!t]
    \centering
    \includegraphics[width=1.0\linewidth]{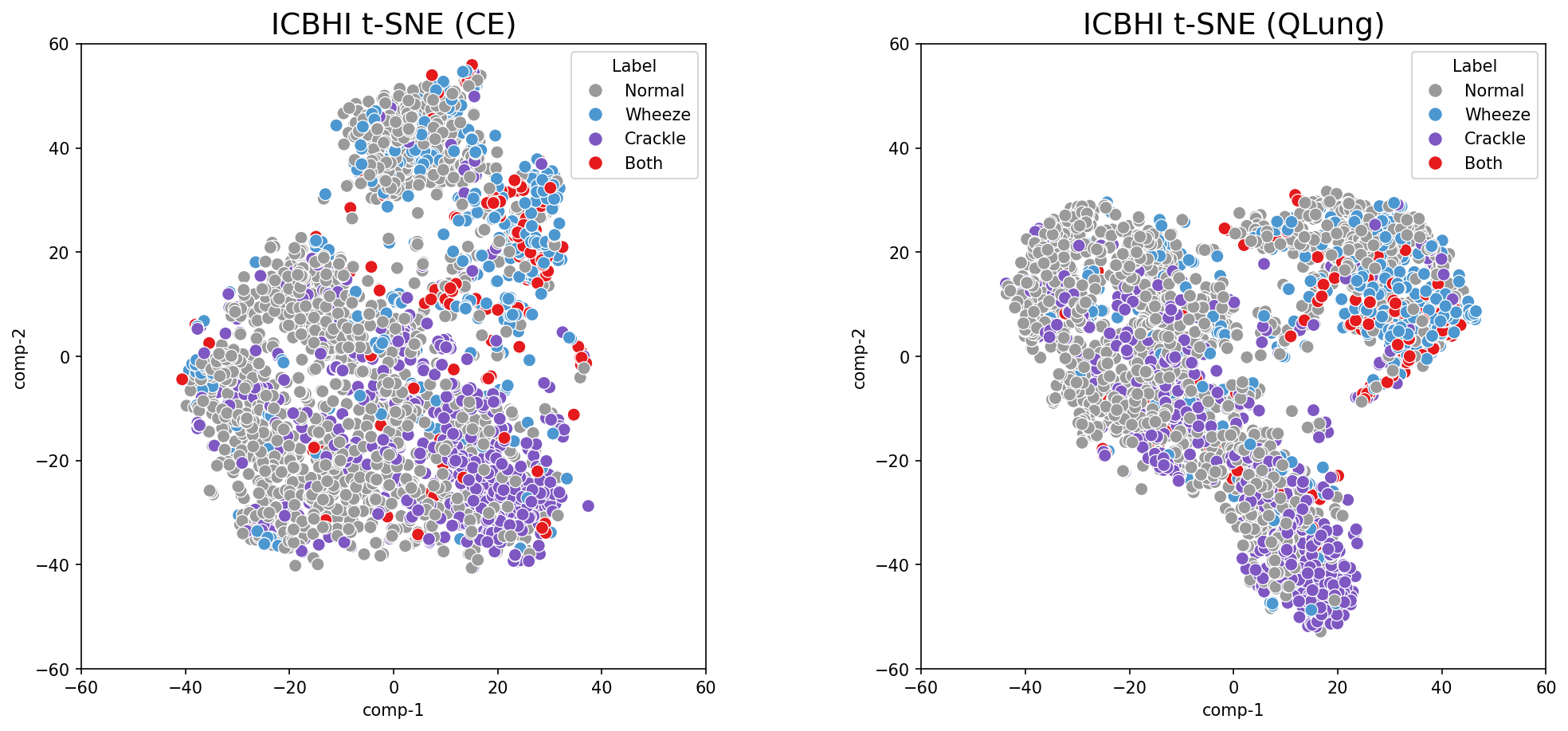}
    \caption{t-SNE visualization of feature embeddings on the ICBHI test set. CE on AST (left) and QLung on AST (right).}
    \label{fig2}    
    \vspace{-5mm}
\end{figure}

\noindent
\textbf{t-SNE Analysis.} 
Figure~\ref{fig2} shows that CE on AST produced dispersed embeddings, with \emph{both} samples scattered across crackle and wheeze clusters. In contrast, QLung yielded more compact and well-separated clusters. Notably, the \emph{both} class formed a distinct cluster, suggesting that QLung learns more discriminative representations for overlapping abnormal events.

\begin{figure}[!t]
    \centering
    \includegraphics[width=1.0\linewidth]{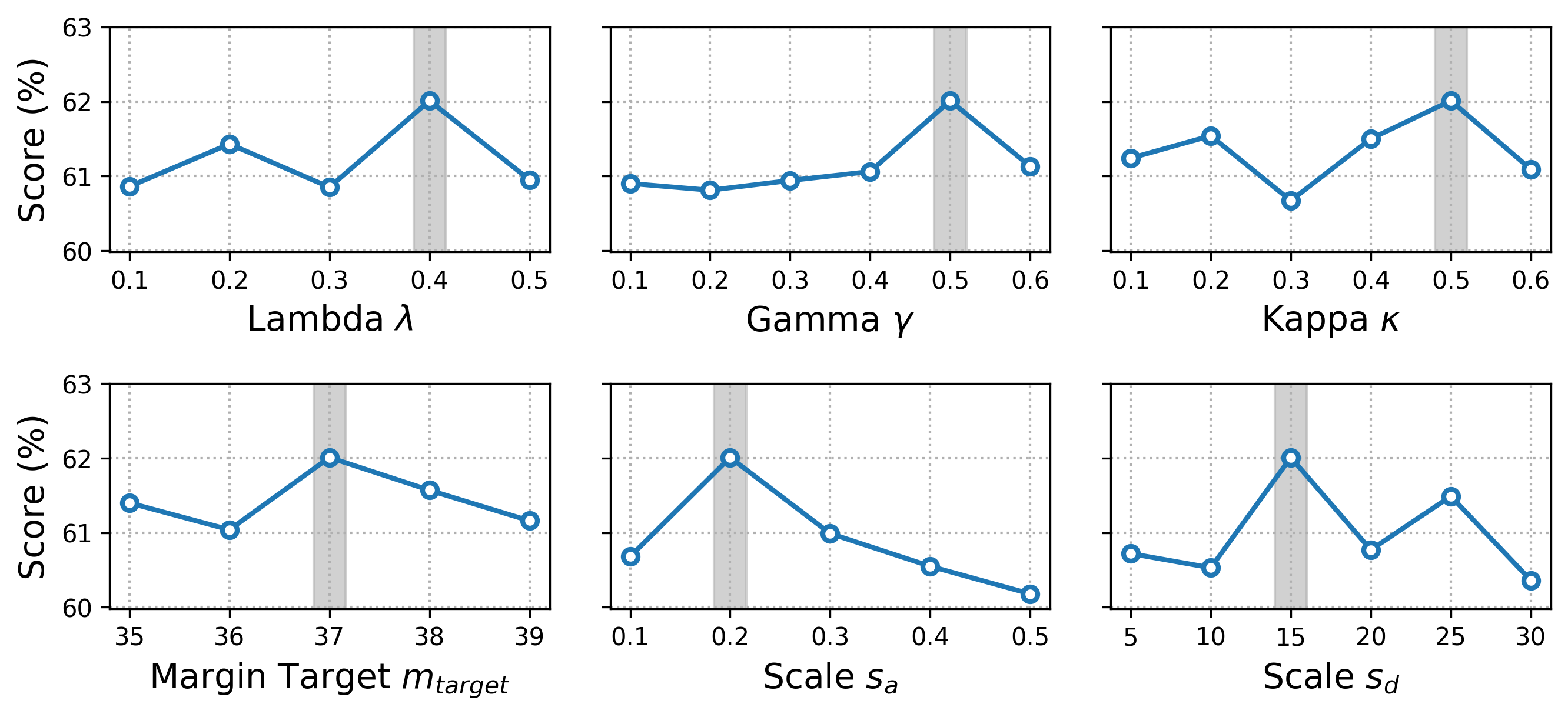}
    \caption{Sensitivity study of QLung hyperparameters. The final configuration is highlighted by the gray shading.}
    \label{fig3}    
    \vspace{-5mm}
\end{figure}

\noindent
\textbf{Hyperparameter Sensitivity Analysis.}
We analyze QLung’s hyperparameter sensitivity in Figure~\ref{fig3}. The Score peaks near the chosen settings and varies only moderately around them, indicating a near-optimal and robust configuration.

\section{Conclusion}
\label{sec:conclusion}
In this paper, we introduced QLung to address the challenges of low-quality recordings and class imbalance in RSC by proposing DFAM with an angular classifier to adapt angular margins. Extensive experiments validated that QLung improved AST fine-tuning by an absolute 2.46\% on the ICBHI dataset and achieved best OOD performance on the SPRSound dataset, indicating that QLung can effectively enhance robustness to noisy data and imbalanced class distributions and highlighting its potential for real-world clinical settings.

\section{Acknowledgement}
This research was supported by the Regional Innovation System \& Education(RISE) program through the Jeonbuk RISE Center, funded by the Ministry of Education(MOE) and the Jeonbuk State, Republic of Korea(2026-RISE-13-WKU), and by the National Research Foundation of Korea(NRF) grant funded by the Korea government(MSIT) (grant no. RS-2025-16066662).

\section{Generative AI Use Disclosure}
Generative AI (ChatGPT) was used solely for grammar correction and linguistic polishing of this manuscript. The authors have verified all technical content and maintain full accountability for the work.

\bibliographystyle{IEEEtran}
\bibliography{mybib}
\end{document}